# Quantum-Inspired Cournot Model

**Amarendra Sharma**[1]

Version 1.0

April 27, 2025

**Abstract**

The primary objective of this paper is to introduce economists to the quantum-inspired approaches of modeling firm behavior in traditional Cournot Duopoly setting in an accessible manner, which could be used for pedagogical purposes. We present quantum concepts such as superposition, entanglement, and quantum search algorithms (Grover's and Dürr-Høyer's) in an intuitive fashion, and show how firms can represent uncertainty, interdependence, and optimization in novel ways. It also incorporates numerical examples and compares outcomes in the classical model with the quantum-inspired outcomes.

## 1. Introduction

The traditional economic models are primarily concerned with the question of how to make decisions under the conditions of scarcity, uncertainty, and interdependence. The concepts of Nash Equilibrium or profit maximization in various market structures, while producing elegant and tractable solutions, assume full rationality and deterministic strategy choices on behalf of economic agents. We draw inspirations from the field of quantum science to explore an alternative framework. In this framework, the firm's decisions are represented by probabilistic amplitudes, and the concept of entanglement is used to represent interdependence. This provides us with a new lens to capture ambiguity, learning, and coordination. It is worth noting that the Quantum inspired models do not require the use of physical Quantum computers to represent decisions. However, they do use mathematical structures from Quantum Mechanics such as Hilbert spaces and state vectors to model complex, adaptive systems. This approach has witnessed growing applications in finance and behavioral science (see for example, Baaquie's (2004); Haven & Khrennikov, 2013). In this paper, we apply the quantum approach to study Cournot quantity competition, which is an important game theoretic model taught in economics courses to model firm behavior in Oligopoly. We aim to provide economists an introduction to these ideas in an accessible manner with real examples and a focus on pedagogy.

[1] Department of Economics, Binghamton University-SUNY, Binghamton, NY, USA. Email: aksharma@binghamton.edu

## 2. Primer on Quantum Concepts for Economists

### 2.1 Superposition

When it comes to decision making in the classical setting, typically an economic agent is assumed to choose one particular action with certainty. Whereas, in quantum mechanics, a system can exist in a superposition. This can be taken as a weighted combination of possible states. Which is to say that economic agents hold a probability distribution across a set of options using amplitudes. For example, suppose a firm is considering choosing quantities 20, 30, or 40 units. This can be represented by:

$$|\psi\rangle = \alpha_0|20\rangle + \alpha_1|30\rangle + \alpha_2|40\rangle$$

Here, $|20\rangle$, $|30\rangle$, and $|40\rangle$, are basis states, and $\alpha_i$ amplitudes such that $|\alpha_0|^2 + |\alpha_1|^2 + |\alpha_2|^2 = 1$. The square of each amplitude represents the probability of observing that particular outcome if the firm is "measured" in quantum sense.

## 2.2 Entanglement

In the classical setting, the firms are assumed to make decisions independently. However, in quantum mechanics, two systems can be entangled. This entanglement property implies that the outcomes of these two systems are correlated beyond what is possible with independent probabilities. In the context of firms, this property allows for strategic interdependence even without explicit communication or expectation formation. Which is to say that two firms can exhibit coordinated behavior in the absence of direct communication. Their choices are correlated through a shared quantum state.

An entangled state can be represented in the following manner:

$$|\psi\rangle = \frac{1}{\sqrt{2}}\left(|30\rangle_1|30\rangle_2 + |40\rangle_1|20\rangle_2\right)$$

The above equation implies that if Firm 1 chooses 30 units to produce, then Firm 2 chooses 30 units as well; and if Firm 1 chooses 40 units, then Firm 2 chooses 20 units. In this representation the alignment between the two firms is built into the shared state and there is no explicit external requirement for either firm to guess the other's strategy. A good example to understand entanglement is the relationship between a lender and a borrower as suggested by Orell (2016). The borrower and lender are linked through a contract and a change in the status of borrower instantaneously affects the status of the lender. In quantum economics, debt is the mechanism that creates an entanglement between the borrower and the lender. This is akin to the property of particles in quantum physics where a change in one particle instantly affects the other.

Let us explore the borrower-lender relationship in the context of a house loan by using a numerical example to illustrate the quantum economics concept of entanglement.

Suppose Alice (the borrower) takes a $200,000 mortgage from Bank B (the lender) to buy a house, with a 30-year term at a fixed 5% annual interest rate. Using compound interest, Alice's monthly payment is approximately $1,073.64, totaling $386,511.60 over 30 years. The house valued at $200,000 at the beginning is subject to depreciation and market fluctuations like any other real-world asset. A financial entanglement has been created by this house loan between Alice and Bank B. Their economic states are now interconnected in a similar manner to the entangled particles in quantum physics, where a change in one particle's state instantly affects the other particles. Bank B's expected cash flow now directly depends on Alice's ability to make payments. Bank B's asset (the house loan) will lose value if Alice is not able to make mortgage payments due to job loss, which could potentially result in foreclosure and Alice being evicted from her house. Conversely, if Bank B increases interest rate on house loan, assuming it's a variable-rate loan, then it will impact Alice's debt burden and will become larger, affecting her financial stability.

### 2.4 Hilbert Spaces and State Vectors

It is common to represent a firm's strategies as either discrete choices or probability distributions in classical economics. However, in quantum-inspired models, a firm's strategies are represented by a state vector in a Hilbert space.

### What is a Hilbert Space?

A Hilbert space is a mathematical space that generalizes Euclidean space. It is a complete vector space equipped with an inner product, which allows for the notions of length and angle. In quantum mechanics, Hilbert spaces provide the environment where all quantum states “live.” For decision-making in economics, one can think of it as a space where each possible strategy (e.g., output level) corresponds to a basis vector.

If a firm has 3 discrete output choices $q \in \{20,30,40\}$, the corresponding Hilbert space is a 3-dimensional complex vector space spanned by the basis vectors:

$$|20\rangle = \begin{bmatrix}1\\0\\0\end{bmatrix}, |30\rangle = \begin{bmatrix}0\\1\\0\end{bmatrix}, |40\rangle = \begin{bmatrix}0\\0\\1\end{bmatrix}$$

**What is a State Vector?**

A state vector (denoted $|\psi\rangle$) is a linear combination of these basis vectors with complex coefficients called amplitudes:

$$|\psi\rangle = \alpha_0|20\rangle + \alpha_1|30\rangle + \alpha_2|40\rangle = \begin{bmatrix}\alpha_0\\\alpha_1\\\alpha_2\end{bmatrix}$$

The amplitudes $\alpha_i$ must satisfy the normalization condition: $|\alpha_0|^2 + |\alpha_1|^2 + |\alpha_2|^2 = 1$.

This ensures that the total probability (when squared) across all possible strategies sums to one.

**Inner Product and Measurement**

The inner product in Hilbert space allows us to compute the expected value of a decision-related operator (e.g., a profit matrix $\widehat{\Pi}$). For a firm in state $|\psi\rangle$, the expected profit is given by: $\langle\psi|\widehat{\Pi}|\psi\rangle$

This is analogous to computing the expected value in probability theory, but it allows for interference and phase relationships that can be exploited algorithmically (e.g., via Grover's algorithm).

**Why Use Hilbert Spaces?**

One of the limitations of classical economics is that firms are required to select one particular strategy (e.g., produce 30 units), or perhaps a mixed strategy (e.g., a probability distribution over options). The advantage of Hilbert Space lies in the notion that a firm's strategic state is a superposition of all possible strategies. Which is to say that a firm can exist in a weighted combination of choices before actually making a decision. For example, suppose a firm considers producing 10, 20, or 30 units of output. In a Hilbert space:

$$|\psi\rangle = 0.5|10\rangle + 0.707|20\rangle + 0.5|30\rangle$$

This implies that the firm has not committed to one particular output but is exploring all possible output options in parallel. The amplitudes here not only encode probabilities, but they also allow for interference effects, which results in some strategies reinforcing or cancelling other strategies. This could be useful in the context of new product launches where the firm is uncertain about the market demand for its product and therefore wants to keep multiple supply plans "in play" while learning the market behavior.

**2.5 Grover's Algorithm**

Grover's algorithm is a quantum search technique that finds a desired item from an unsorted list in $\sqrt{n}$ steps rather than in $n$ steps. If we apply it to our context in economics, then it can help a firm in identifying the most profitable output from many output options in an efficient manner. This algorithm relies on amplitude amplification to efficiently search for the best output that maximizes profit. In the classical world, searching for the best output over many possibilities is time consuming because the search process is linear in the number of options. The Grover's algorithm exploits the advantage afforded by Hilbert Space and enables the firm to speed up the search process in a quadratic fashion. It uses structured amplitude amplification to focus the quantum state on the optimal strategy. For example, if a firm has 100 possible output choices, then the classical search will take 100 steps to solve it. Whereas the Grover's algorithm will need only ~10 steps. One can think of Grover's algorithm as an intelligent trial-and-error process, where the search space "resonates" toward the correct answer, rather than blindly checking each one. In volatile markets, this quadratic search speed could offer a significant advantage.

### 2.6 Dürr-Høyer Algorithm

This algorithm is an extension of Grover's algorithm and finds the global minimum in unknown distributions. It closely mimics the behavior of real firms. We consider a variant of this algorithm (see for example, Baritompa et.al, 2005). It starts with a guess about the quantity and then adapts if better options exist. The advantage of using this algorithm lies in its efficiency in terms of time saved in comparison to brute-force checking, especially for the larger strategy spaces. In our context, this algorithm is particularly advantageous when the maximum profit is an unknown quantity, unlike the Grover's algorithm that relies on marking the maximum profit before commencing the search process.

## 3. Literature Review

Quantum concepts have progressively influenced economic modeling by providing novel insights into decision-making, strategic behavior, and market equilibrium. Asghar Qadir, a Pakistani mathematician and physicist, is broadly acknowledged as the first person to directly relate and apply quantum tools to economic analysis. In 1978, he published a seminal paper titled *Quantum Economics*, which laid the foundation for the field. Qadir suggested that the framework of quantum mechanics could be used to model complex consumer behavior, since consumer preferences depend on myriad factors and are not fully known until the consumer makes the decision. He proposed treating individuals as entities in a Hilbert space, akin to particles in quantum mechanics, to more aptly represent the uncertainty and interconnectedness of economic decisions.

Baaquie's (2004) foundational work laid the groundwork for quantum finance by incorporating quantum mechanics concepts into financial theory. Building on this work, Haven and Khrennikov (2013) formalized quantum-like modeling in the social sciences. Their work postulated that cognitive and economic phenomena might be better represented within the probabilistic structure of quantum theory than the existing classical models.

In the domain of game theory, Meyer (1999) was the first researcher to demonstrate that quantum strategies could outperform classical strategies in simple games, thereby highlighting the strategic advantage of quantum superposition. Iqbal and Toor (2002) extended these ideas to Prisoners' dilemma game. They showed that the Nash equilibria in these games could be altered due to quantum entanglement of strategies.

The formal quantum version of the Cournot duopoly game was proposed by Li, Du, and Massar (2002). In their version the firms use entangled strategies. They show that the quantum entanglement can yield cooperative behavior that is unachievable in classical Nash equilibria.

On the Cognitive side, the work of Pothos and Busemeyer (2009) introduced a quantum cognitive framework for decision-making under uncertainty. Their model explains paradoxes like sure thing principle of decision theory which challenge classical decision theory.

Recent work by Yukalov & Sornette (2009) show that the quantum approach avoids any paradox typical of classical decision theory.

Bagarello (2021) discusses how quantum mechanics, which traditionally has been associated with microscopic systems, can however also be applied to a diverse range of macroscopic systems both within and outside of physics. This work describes the mechanisms to model complex systems from a variety of fields, such as biology, ecology, decision-making, and sociology, using quantum mechanical principles

Our work contributes to this body of literature by intoducing Grover's search algorithm and Dürr-Høyer's optimization algorithm to simulate optimal firm behavior in Cournot duopoly setting. We develop accessible simulations and numerical examples to make the tools of quantum-inspired modeling pedagogically relevant and computationally tractable for economics education. Further, our use of entangled strategies as a device in lieu of the need for explicit expectation formation by firms in classical settings is a novel contribution to firm theory. Finally, we numerically show how such quantum models can improve on classical Pareto efficiency in certain strategic environments.

## 5. Classical Versus Quantum Cournot Model

In this section, we compare the classical Cournot competition with the Quantum version of the model. We consider a simple case of duopoly where the two firms produce identical goods and have access to same technology to produce the good resulting in same cost of production.

### 5.1 Classical Cournot Model

Two firms choose output simultaneously. Let demand be:

$$P(Q) = 100 - Q = 100 - (q_1 + q_2)$$

Costs: $C(q) = 10q$

Profit: $\Pi_i = (100 - q_1 - q_2)q_i - 10q_i$

Best response or Reaction function for Firm 1:

$$q_1^* = \frac{90 - q_2}{2}$$

Symmetric Nash Equilibrium:

$$q_1^* = q_2^* = 30, \Pi_1 = \Pi_2 = 900$$

### 5.2 Quantum-Inspired Cournot Model

Let us suppose that firms choose quantum states over outputs ({20, 30, 40}) and jointly agree on the following entangled quantum state (strategy):

$$|\Psi\rangle = \frac{1}{\sqrt{2}}(|30\rangle|30\rangle + |40\rangle|20\rangle)$$

Each outcome occurs with 50-percent probability.

$$\text{Case 1: } (q_1, q_2) = (30,30),\ \Pi_1 = \Pi_2 = 900$$

$$\text{Case 2: } (q_1, q_2) = (40,20),\ \Pi_1 = (100 - 60)(40) - (10)(40) = 1200$$

$$\Pi_2 = (100 - 60)(20) - (10)(20) = 600$$

Next, compute expected profits:

$$\mathrm{E}[\Pi_1] = \frac{1}{2}(900 + 1200) = 1050$$

$$\mathrm{E}[\Pi_2] = \frac{1}{2}(900 + 600) = 750$$

The advantage of entanglement is that the firms are not required to form expectations. The entanglement embeds strategic alignment.

**6. Strategy Optimization via Grover's Algorithm**

Suppose firms choose from four discrete output levels {10, 20, 30, 40}. This implies that there are $(4)^2 = 16$ possible quantity pairs that will need to be evaluated in classical optimization. Grover's algorithm requires only $\sqrt{16} \approx 4$ queries to find the maximum.

For simplicity, let's just consider a sample of the quantity pairs $[(10,10), (20,20), (30,30), (40,40)]$ and the associated total profits of the industry consisting of the two firms

$$\Pi(\mathrm{Q}) = [1400, 2000, 1800, 800]$$

Using Grover's oracle and diffusion steps, we initialize with equal amplitudes. The oracle marks Q = 40 (max profit = 2000). After 2 iterations: $|40\rangle$ dominates with $\approx$90% measurement probability.

Grover's method scales efficiently and allows firms to explore large strategy spaces without exhaustive search (Please refer to Appendix A for details). Interestingly, in this example, Grover's search algorithm finds the outcome where the two firms choose 20 units each for a total of 40 units. This outcome is similar to the classical case of Cournot duopoly with collusion where the two firms jointly decide to produce a monopoly output and split it equally.[2] However, here the collusion is not explicitly required because of the entanglement.

## 6. Strategy Optimization via Dürr-Høyer's Algorithm

To demonstrate optimization using Dürr-Høyer quantum minimum/maximum search algorithm, we use the same Cournot example as in the previous section. Here we'll assume that the firm doesn't know the optimum value of profit in advance. For illustrative purpose, let's consider a subset of the output space to compute profit by fixing the quantity of firm 2 to 20 units. The table below presents the profits for each output choice. This is our classical baseline.

**Table 1**

| $q_1$ | Total Output Q | Price P | Revenue | Cost | Profit |
|---|---|---|---|---|---|
| 10 | 30 | 70 | 700 | 100 | 600 |
| 20 | 40 | 60 | 1200 | 200 | 1000 |
| 30 | 50 | 50 | 1500 | 300 | 1200 |
| 40 | 60 | 40 | 1600 | 400 | 1200 |

[2] Similar result has been shown in a formal setting by Li, Du, and Massar (2002).

It is evident that the best profit is 1200, occurring at both $q_1$=30 and $q_1$=40. Next, let's consider the intuition behind the workings of Dürr-Høyer algorithm.

1. Randomly guess a solution $q_1$ — call it the "current best."
2. Use quantum search (like Grover's) to look for any better value.
3. If a better value is found, replace the current best, and repeat.
4. After a number of rounds (proportional to √N), the probability of having found the maximum is high.

This algorithm differs from Grover's, which assumes that we know the target (oracle marks it). Dürr-Høyer's algorithm searches adaptively for the target and is ideal for cases when one doesn't know the optimum value in advance. Next, we apply Dürr-Høyer to our problem.

Iteration 1: Pick a random starting value. Say we pick $q_1$=20, profit = 1000. Next, use Grover-style search to find any value with profit > 1000. The algorithm will find $q_1$=30,40. Both outputs have profit of 1200. The algorithm will update the best value to $q_1$=30.

Iteration 2: The current best value is 1200. Use Grover-like search again to find any value > 1200. The algorithm doesn't find any such values. The algorithm will terminate with $q_1$ = 30 or 40, profit = 1200.

**7. Pareto Optimality and Welfare Comparisons**

To evaluate whether quantum-inspired models offer not only computational or strategic benefits but also welfare improvements, we compare the classical and quantum outcomes. The benchmark is Pareto optimality—whether the quantum outcome makes at least one party better off without making others worse off.

**Table 2: Comparative Outcomes**

| Market Structure | Classical Output | Classical Profit (Firm) | Classical Consumer Surplus | Quantum Output | Quantum Profit (Firm) | Quantum Consumer Surplus |
|---|---|---|---|---|---|---|
| Cournot Oligopoly | (30,30) | (900,900) | (600, 600) | (20,20) | (1000,1000) | (600, 600) |

The surplus consisting of consumer surplus and producer surplus is higher under the quantum case. Producer surplus is higher relative to the classical case and the consumer surplus is the same. This shows that quantum firms can choose an output pair without coordination that is pareto improving (Please refer to Appendix B for calculations pertaining to Pareto optimality and Welfare comparison).

## 8. Conclusion

In this paper we discuss intuitive tools inspired by Quantum principles to understand firm behavior under uncertainty and interdependence. We use simplified quantum computation analogies and numerical examples to explain coordination, optimization, and decision-making in complex economic environments.

We show that enriched forms of coordination (via entanglement) and efficient optimization (via Grover and Dürr-Høyer algorithms) can be accomplished in the quantum-inspired models. In oligopolistic setting, these features could translate into outcomes that are both privately optimal and socially beneficial.

Through our analysis, we highlight that quantum-inspired decision framework can yield Pareto superior outcomes in certain cases. It gives us a new lens for examining market efficiency and strategic interdependence. The advantage of this framework lies in the use of entangled strategies that allows firms to explore coordinated quantity choices, even in non-cooperative games. This enables firms to sample joint strategies rather than focus only on unilateral ones. Firms can escape local optima (like the Nash equilibrium) and move to mutually beneficial outcomes by fine-tuning the amplitudes of entangled outcomes (analogous to coordination probabilities). This scenario is impossible in classical Cournot games without explicit cooperation, but quantum-inspired entanglement provides a non-cooperative channel to coordination.

## Appendix A

### Grover's Algorithm

Find the quantity q ∈ {10, 20, 30, 40, 50, 60, 70,80} that gives the highest profit using Grover's algorithm.

The associated Profits are: Π(q) = [500,800,1200,2000,1700,1400,1000,600]

Classically, we will need to evaluate all 8 quantity entries. Grover's algorithm performs it in about $\sqrt{8} \approx 2.8 \approx 3$ steps.

**Step 1:** Represent the Search Space in a Quantum State

We start by encoding all 8 strategies (quantities) as quantum basis states:

$|q\rangle \in \{|0\rangle, |1\rangle, |2\rangle, |3\rangle, |4\rangle, |5\rangle, |6\rangle, |7\rangle\}$

Let's label them:

$|0\rangle \equiv q=10$, $|1\rangle \equiv q=20$, $|2\rangle \equiv q=30$, $|3\rangle \equiv q=40$ (the correct answer), $|4\rangle \equiv q=50$, $|5\rangle \equiv q=60$, $|6\rangle \equiv q=70$, $|7\rangle \equiv q=80$

Then initialize the quantum state to a uniform superposition over all 8 options:

$$|\psi_0\rangle = \frac{1}{\sqrt{8}}\sum_{i=0}^{7}|i\rangle$$

This means the firm is equally considering all 8 output levels.

**Step 2:** Grover Oracle – Mark the correct answer

Now we apply an oracle—a black-box quantum operation—that flips the sign of the amplitude of the correct answer (the quantity with maximum profit).

Let's assume the oracle "knows" that q=40 (i.e.,$|3\rangle$) gives the highest profit. It flips the sign of that amplitude:

Before oracle: $|\psi\rangle=\sqrt{\frac{1}{\sqrt{8}}}[1,1,1,1,1,1,1,1]$. After oracle: $|\psi\rangle=\sqrt{\frac{1}{\sqrt{8}}}[1,1,1,-1,1,1,1,1]$

Now only the correct answer (state 3) has a negative amplitude.

**Step 3:** Diffusion Operator – Amplify the Marked Answer

Now we apply a special operator called the diffusion operator, which reflects each amplitude about the average. This step is what amplifies the probability of the marked state (here, $|3\rangle$).

Reflection Around the Average: The Rule

Suppose the state of the system after the oracle is:

$$|\psi\rangle = [a_0, \ldots\ldots, a_k]$$

The Grover diffusion operator does the following reflection:

$$a_i \rightarrow 2\overline{a} - a_i$$

Where:

$a_i$is the amplitude of basis state *i*

$\overline{a}$ is the average of all amplitudes

This is literally a reflection: if a value is above the average, it gets pushed below, and vice versa. It preserves the symmetry and amplifies the deviations.

Suppose we have the post-oracle state for 8 options:

$$a = \left[\frac{1}{\sqrt{8}}, \frac{1}{\sqrt{8}}, \frac{1}{\sqrt{8}}, -\frac{1}{\sqrt{8}}, \frac{1}{\sqrt{8}}, \frac{1}{\sqrt{8}}, \frac{1}{\sqrt{8}}, \frac{1}{\sqrt{8}}\right]$$

That's approximately:

$$a \approx [0.3535, 0.3535, 0.3535, -0.3535, 0.3535, 0.3535, 0.3535, 0.3535]$$

Next, compute the average

$$\overline{a} = \frac{7(0.3535) + (-0.3535)}{8} = \frac{2.47458}{8} \approx 0.3093$$

Apply reflection:

$$a_i \rightarrow 2\overline{a} - a_i$$

$a_0 = 0.3535$

$$a_0' = 2(0.3093) - 0.3535 = 0.6186 - 0.3535 = 0.2651$$

$a_3 = 0.3535$

$a_3' = 2(0.3093) - (-0.3535) = 0.6186 + 0.3535 = 0.9721$

So, the amplitude of the marked state (the one the oracle flipped) grows significantly, while the others shrink slightly. This shows how Grover's algorithm amplifies the probability of the optimal answer.

**Step 4:** Measurement

The firm "measures" the state—that is, it chooses an action. Since $|3\rangle$ (i.e., q=40) dominates the state, it will almost certainly be selected.

# Appendix B

Below we present Excel spreadsheet calculations for Welfare comparisons. The demand and cost conditions are from the original model.

| q1 | q2 | Q | P | P.q1 | tc1 | Pi1 | p.q2 | tc2 | Pi2 | total Pi | amp | oracle | diff | NEW AMP | DIFF 2 | NEW AMP2 | CS | W |
|---|---|---|---|---|---|---|---|---|---|---|---|---|---|---|---|---|---|---|
| 10 | 10 | 20 | 80 | 800 | 100 | 700 | 800 | 100 | 700 | 1400 | 0.25 | 1 | 0.25 | 0.1875 | 0.1875 | 0.078125 | 800 | 2200 |
| 10 | 20 | 30 | 70 | 700 | 100 | 600 | 1400 | 200 | 1200 | 1800 | 0.25 | 1 | 0.25 | 0.1875 | 0.1875 | 0.078125 | 1050 | 2850 |
| 10 | 30 | 40 | 60 | 600 | 100 | 500 | 1800 | 300 | 1500 | 2000 | 0.25 | 1 | 0.25 | 0.1875 | 0.1875 | 0.078125 | 1200 | 3200 |
| 10 | 40 | 50 | 50 | 500 | 100 | 400 | 2000 | 400 | 1600 | 2000 | 0.25 | 1 | 0.25 | 0.1875 | 0.1875 | 0.078125 | 1250 | 3250 |
| 20 | 10 | 30 | 70 | 1400 | 200 | 1200 | 700 | 100 | 600 | 1800 | 0.25 | 1 | 0.25 | 0.1875 | 0.1875 | 0.078125 | 1050 | 2850 |
| 20 | 20 | 40 | 60 | 1200 | 200 | 1000 | 1200 | 200 | 1000 | 2000 | 0.25 | -1 | -0.25 | 0.6875 | -0.688 | 0.953125 | 1200 | 3200 |
| 20 | 30 | 50 | 50 | 1000 | 200 | 800 | 1500 | 300 | 1200 | 2000 | 0.25 | 1 | 0.25 | 0.1875 | 0.1875 | 0.078125 | 1250 | 3250 |
| 20 | 40 | 60 | 40 | 800 | 200 | 600 | 1600 | 400 | 1200 | 1800 | 0.25 | 1 | 0.25 | 0.1875 | 0.1875 | 0.078125 | 1200 | 3000 |
| 30 | 10 | 40 | 60 | 1800 | 300 | 1500 | 600 | 100 | 500 | 2000 | 0.25 | 1 | 0.25 | 0.1875 | 0.1875 | 0.078125 | 1200 | 3200 |
| 30 | 20 | 50 | 50 | 1500 | 300 | 1200 | 1000 | 200 | 800 | 2000 | 0.25 | 1 | 0.25 | 0.1875 | 0.1875 | 0.078125 | 1250 | 3250 |
| 30 | 30 | 60 | 40 | 1200 | 300 | 900 | 1200 | 300 | 900 | 1800 | 0.25 | 1 | 0.25 | 0.1875 | 0.1875 | 0.078125 | 1200 | 3000 |
| 30 | 40 | 70 | 30 | 900 | 300 | 600 | 1200 | 400 | 800 | 1400 | 0.25 | 1 | 0.25 | 0.1875 | 0.1875 | 0.078125 | 1050 | 2450 |
| 40 | 10 | 50 | 50 | 2000 | 400 | 1600 | 500 | 100 | 400 | 2000 | 0.25 | 1 | 0.25 | 0.1875 | 0.1875 | 0.078125 | 1250 | 3250 |
| 40 | 20 | 60 | 40 | 1600 | 400 | 1200 | 800 | 200 | 600 | 1800 | 0.25 | 1 | 0.25 | 0.1875 | 0.1875 | 0.078125 | 1200 | 3000 |
| 40 | 30 | 70 | 30 | 1200 | 400 | 800 | 900 | 300 | 600 | 1400 | 0.25 | 1 | 0.25 | 0.1875 | 0.1875 | 0.078125 | 1050 | 2450 |
| 40 | 40 | 80 | 20 | 800 | 400 | 400 | 800 | 400 | 400 | 800 | 0.25 | 1 | 0.25 | 0.1875 | 0.1875 | 0.078125 | 800 | 1600 |

Where q1=output choice of firm 1, q2= output choice of firm 2, Q= total industry output, P.q1=Revenue of firm 1, tc1=total cost of firm 1, Pi1=profit of firm 1, P.q2=Revenue of firm 2, tc2=total cost of firm 2, Pi2=profit of firm 2, total Pi=Total Industry profit, amp= initial amplitude set at equal values for all possible states, oracle= oracle marks the highest profit assuming symmetric output choices considering both firms are identical, diff=diffusion, NEW AMP=New amplitudes after first reflection, DIFF2= amplitudes after second diffusion, NEW AMP2= updated amplitudes after second reflection, CS=Consumer Surplus, W=Welfare